\documentclass[prb,twocolumn]{revtex4-1} % style for Physical Review B and AJP are similar

\usepackage{comment}
\usepackage{amsmath} % need for subequations
\usepackage{amssymb}

\usepackage{graphicx} % for figures

\usepackage{hyperref}
\hypersetup{colorlinks=true,linkcolor=red,citecolor=blue,urlcolor=black,bookmarksdepth=subsubsection} %http://en.wikibooks.org/wiki/LaTeX/Hyperlinks#cite_note-1

\begin{document}

\title{Artificial circumzenithal and circumhorizontal arcs}
\author{Markus Selmke and Sarah Selmke*}
\affiliation{*Universit\"at Leipzig, 04103 Leipzig, Germany}
\email{markus.selmke@gmx.de}
\homepage{http://photonicsdesign.jimdo.com}
\date{\today}

\begin{abstract}
We revisit a water glass experiment often used to demonstrate a rainbow. On a closer look, it also turns out to be a rather close analogy of a different kind of atmospheric optics phenomenon altogether: The geometry may be used to faithfully reproduce the circumzenithal and the circumhorizontal halos, providing a missing practical demonstration experiment for those beautiful and common natural ice halo displays.
\end{abstract}

%e.g.: https://opticalillusion.wordpress.com/2008/07/21/how-to-make-a-rainbow-at-home/
%http://lifestyle.iloveindia.com/lounge/how-to-make-a-rainbow-2613.html
%http://www.asu.edu/courses/phs208/patternsbb/PiN/mod/light/opticsnature/pattLight4Obj1.html
%http://www.nws.noaa.gov/os/educ/activit/rainbow.htm
%http://redtri.com/how-to-make-a-rainbow/
%http://www.scienceprojectideas.co.uk/making-rainbow-breaking-light-into-colour.html
%http://www.kidsgen.com/school_projects/make_rainbow.htm
%http://www.arvindguptatoys.com/arvindgupta/physicsexperiments.pdf
%http://www.madsci.org/posts/archives/1999-01/916547183.Ch.r.html
%"Gilbert light experiments for boys", Experiment No. 94 "Artificial Rainbow", The A.C. Gilbert Company: https://archive.org/details/gilbertlightexpe00lynd
%"Early Education Curriculum: A ChildÕs Connection to the World", Hilda Jackman, p. 274, https://books.google.de/books?id=h5bab-0_5YUC&pg=PA274&lpg=PA274&dq=glass+water+table+rainbow&source=bl&ots=JG65u6Wo3G&sig=85m_0hftJNRFq1qHmZX4XQAgBbA&hl=de&sa=X&ved=0ahUKEwjgj6Pd9-nOAhWGPRoKHbt-AlEQ6AEIbzAN#v=onepage&q=glass%20water%20table%20rainbow&f=false,
%
\maketitle %KERN ARC??, http://iapetus.jb.man.ac.uk/cza/CZA.html

%"Applied Optics", Ed E. Miller and Fred L. Roesler, https://www.physics.wisc.edu/undergrads/courses/spring10/625/Miller&Roesler/Miller-Roesler%20Book%20Figures/Book%20Tex%20Info/Miller-Roesler_Book%20folder/Active%20Assembling%20folder/AAAOpticsRootBook.pdf
%page 126, top: "Nowadays I demonstrate this and several other ice crystal displays using a water-filled, hexagonal box of microscope slides epoxyed together with a hexagonal end plate of glass.Ó
%"Laboratory experiments in atmospheric optics", Michael Vollmer and Robert Tammer: "Two of the most bril- liant arcs, the circumzenithal and the circumhorizon- tal arcs  due to crystals with c axes vertical and a refracting angle of 90¡ , are, unfortunately, observ- able only for crystals with an index of refraction of n    2. Hence they are not observable with the glass prisms and hexagons. It is, however, possible, to observe these arcs by use of a hollow hexagon  made of microscope slides  filled with water."

\section{Introduction}
Light which falls onto a transparent thin-walled cylinder (e.g.\ a drinking glass) filled with water gets refracted. Several ray paths may be realized through what then effectively represents a cylinder made of water. Light may either illuminate and enter through the side of the cylinder, or may enter through the top or bottom interfaces, depending on the angle and spot of illumination. Indeed, in the former situation, i.e.\ illumination from the side and under a shallow inclination angle reveals a rainbow in the backwards direction. The reason being that the geometry mimics the incidence plane geometry of a light path though a spherical raindrop: Refraction, internal reflection and a second refraction upon exit, \textit{all occurring at the cylinder's side wall}, produce the familiar observable rainbow caustic in the backwards direction at around $42^{\circ}$ towards the incidence light source.\cite{Greenler1980,Casini2012} %The mechanism responsible for the increased intensity is the bundleling of rays close to the minimum deflection angle, an effect known as a caustic.

\begin{figure}[t]
\begin{center}\includegraphics [width=1.0\columnwidth]{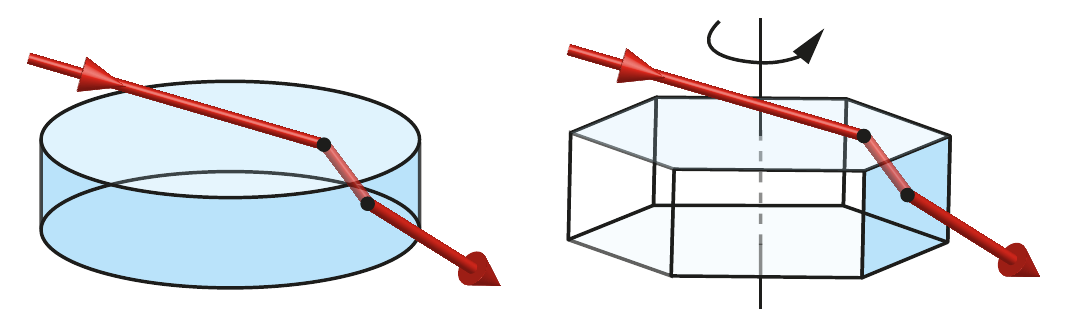}\end{center}
\caption{Rays entering through the top face of both a cylinder (left) and a hexagonal prism (right) experience an equivalent refraction. Refraction of the skew rays by the side faces are equivalent when the effect of rotational averaging of the prism is considered. The same holds true for the reverse ray path.\label{Fig_Corresp}}
\end{figure}

Now, returning to the initial claim, we consider illumination of the glass \textit{through the top water-air interface}. If the angle of incidence is shallow enough, light may \textit{exit through the cylinder's side wall}. Contrary to common belief,\cite{Gilbert1920} (cf.\ also blogs etc.\ found via an internet search for \lq\lq glass water table rainbow\rq\rq) this situation is \textit{not} related to the rainbow. Instead, this geometry equals the \textit{average} geometry of light paths through an upright hexagonal ice prism, entering through the (horizontal) top face and leaving through either of its six (vertical) side faces, cf.\ Fig.\ \ref{Fig_Corresp}. The averaging meant being over different prism orientations as indicated in the figure. This, in turn, is what causes the natural atmospheric phenomenon known as the circumzenithal arc (CZA) halo,\cite{Bravais1847,Humphreys,Ticker,Wegener,McDowell1979,Greenler1980,TapeBook,TapeJarmo2006,atmosHP}  an example of which is shown in Fig.\ \ref{Fig_Exp}(a). In the experiment, an analogous curved spectrum is observed when the refracted light is projected on the floor (the horizontal plane) some distance from the cylinder,\cite{Distance} see Fig.\ \ref{Fig_Exp}(b).

Similarly, illuminating the glass at a very steep angle at its side, the light may \textit{enter through the side wall and leave through the top surface}. Now, apart from top and bottom being reversed, this geometry equals the \textit{average} geometry of light entering a rectangular (vertical) side face of a hexagonal plate crystal and leaving through its bottom (horizontal) hexagonal face. This is the situation corresponding to the natural halo phenomenon known as the circumhorizontal arc (CHA).\cite{Bravais1847,Humphreys,Ticker,Wegener,Greenler1980,TapeBook,TapeJarmo2006,atmosHP} 

Anecdotally, it appears puzzling why Huygens, who was the first to establish an extensive quantitative framework for halos based on the (false) assumption of refracting cylinders, did not conceive of this CZA mechanism and instead invoked a more complicated one.\cite{Puzzle}

We will detail each experimental setup and show how to arrive at a quantitative description of several aspects of the artificial halo analoga, rederiving well-known expressions from the natural atmospheric optics ice halo phenomena. For ideal experimental results, one may use a round reflection cuvette. However, a beaker or any other cylindrical glass and a focusable LED flashlight (source of parallel white light) will work just fine.

\section{Artificial circumzenithal arc}
\begin{figure*}[hbt!]
\begin{center}\includegraphics [width=1.0\textwidth]{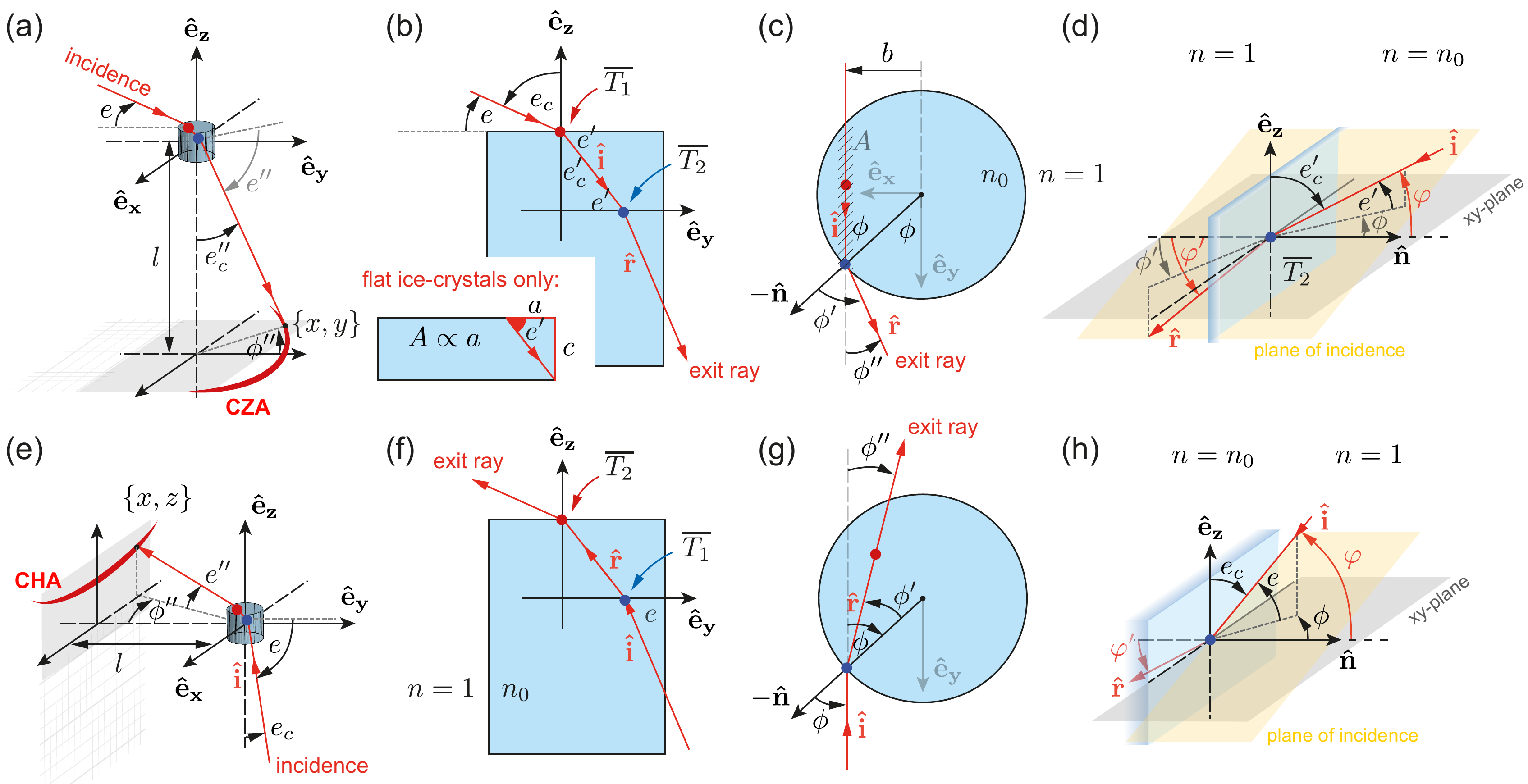}\end{center}
\caption{Geometry and setup for \textbf{(a)-(d)} the CZA experiment and \textbf{(e)-(h)} the CHA experiment. Read text for details.\label{Fig_Geometry}}
\end{figure*}

We begin with the artificial CZA, for which a ray is assumed to enter through the top air-water interface and to leave the cylinder through its side wall, cf.\ Fig.\ \ref{Fig_Geometry}(a)-(c). At the first interface, the ray changes its inclination $e$ towards the horizontal plane according to Snell's law. We denote complementary angles by a subscript $c$, such that for instance $e_c=\pi/2 - e$, see Fig.\ \ref{Fig_Geometry}(b). Thus, we have $\sin\left(e_c\right)=n_0\sin\left(e_c'\right)$, with an associated transmission coefficient $T_1\left(e_c\right)$ according to the Fresnel equations. When later discussing intensities, we will consider polarization-averaged transmission coefficients only, although this approach will not strictly be valid for the second refraction due to the partial polarization upon the first refraction. 

The second (skew-ray) refraction now occurs under a geometry that may be decomposed into two parts\cite{Bravais1847,Humphreys,Ticker,Wegener,Koennen1983,Selmke2015}: One in the horizontal plane (i.e.\ as seen from the top, cf.\ Fig.\ \ref{Fig_Geometry}(c)) and described by an effective index of refraction $n'$, Bravais' index of refraction for inclined rays, and a second inclination refraction described by the actual material's index of refraction $n_0$. The appropriate effective index of refraction reads:\cite{neff}
\begin{equation}
n'=\sqrt{\frac{\cos^2\left(e'\right)}{1/n_0^2 - \sin^2\left(e'\right)}}.\label{eq:neff}
\end{equation}
\begin{figure}[h!]
\begin{center}\includegraphics [width=1.0\columnwidth]{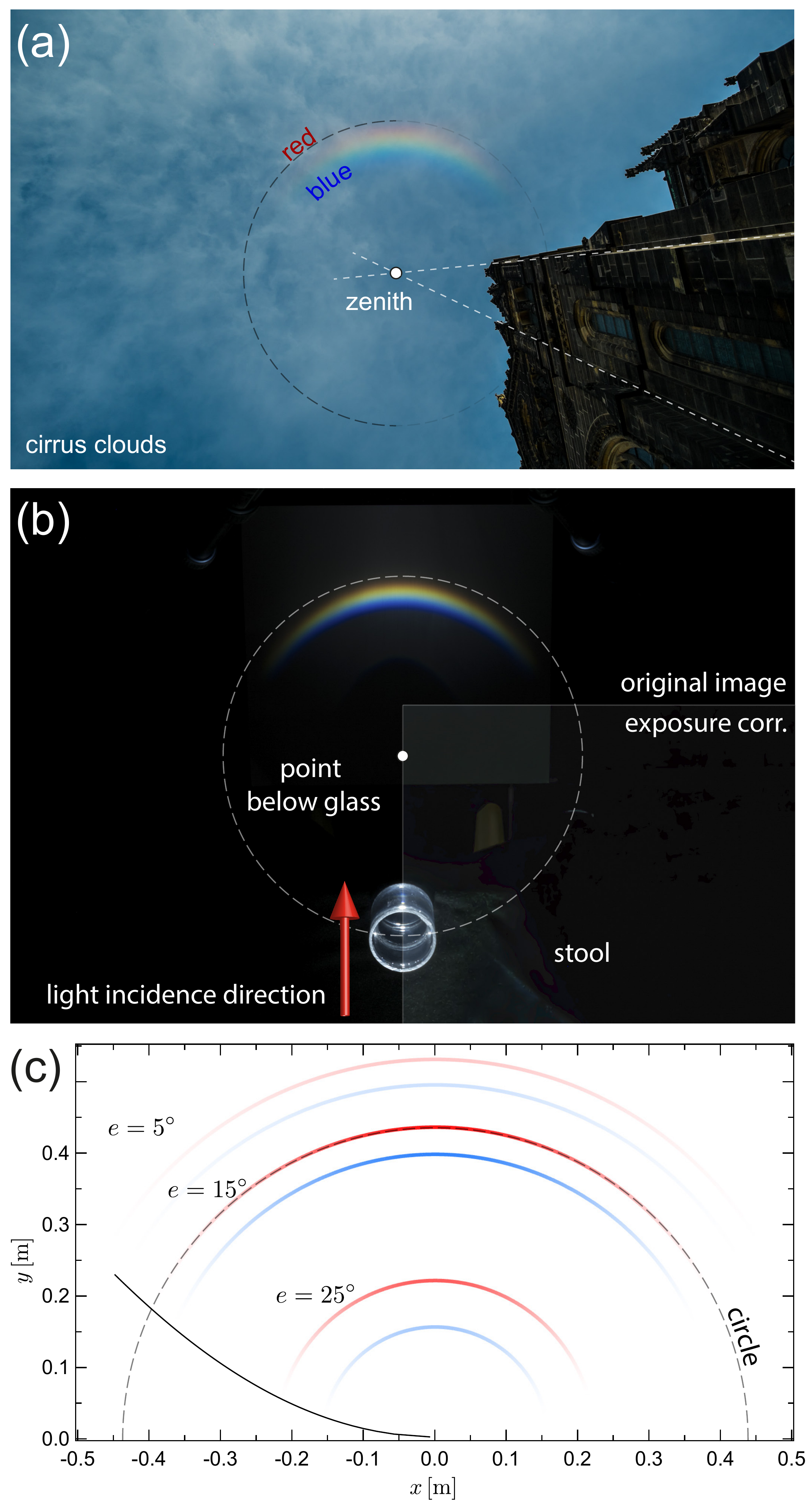}\end{center}
\caption{\textbf{(a)} Natural circumzenithal arc (CZA) halo display. The solar elevation was $e=27^{\circ}$,\cite{Wolfram} and the angular distance to the zenith was $e''_c=16^{\circ}$ as determined with the help of the vanishing lines of vertical \& parallel features of the building (dashed lines).\cite{CZAnat} \textbf{(b)} Artificial CZA produced by illuminating the top surface of a water-filled acrylic cylinder under a shallow angle (outer $\varnothing=50\,\rm mm$, inner $\varnothing=46\,\rm mm$, length: $50\,\rm mm$, height: $l=1\,\rm m$). \textbf{(c)} Artificial CZA curves according to Eq.\ \eqref{eq:CZA} for $e=\left\{5^{\circ},15^{\circ},25^{\circ}\right\}$ and for red and blue color each. Shading according to the intensity $I\left(\phi''\right)$, Eq.\ \eqref{Eq:Int}.\label{Fig_Exp}}
\end{figure}
The exiting ray, which hit the cylinder's side wall under a $xy$-projected incidence angle of $\phi$ (to the normal), is thus deflected in the horizontal plane by $\phi''=\phi'-\phi$, where Snell's law connects the latter two angles via $n'\sin\left(\phi\right)=\sin\left(\phi'\right)$. The inclination angle to the plane changes according to $n_0\sin\left(e'\right)=\sin\left(e''\right)$, such that overall the exit angle to the vertical becomes\cite{Bravais1847,Humphreys,Wegener,McDowell1979} 
%Bravais: p.30, p. 96, p. 233 (Note VI, 5e Cas, (47))? l=n0, alpha=A (prisms' apex angle), H=e (elevation), h=e''.
%
\begin{equation}
e''_c= \arccos\left(\sqrt{n_0^2-\cos^2\left(e\right)}\right)\label{Eq:eppc}.
\end{equation}
Then, referring to the experiment's setup and coordinates as defined in Fig.\ \ref{Fig_Geometry}(a), one finds for each light source inclination angle $e$ the deflected rays to lie on a curve $\left(x\left(\phi\right),y\left(\phi\right)\right)$. This CZA curve may be parametrized by the angle $\phi\in \left[-\pi/2,\pi/2\right]$, see Fig.\ \ref{Fig_Geometry}(c), as
\begin{equation}
\left(\begin{array}{c}x\\y\end{array}\right)=l\tan\left(e''_c\right)\left(\begin{array}{c}\sin\left(\phi''\right)\\\cos\left(\phi''\right)\end{array}\right)\label{eq:CZA},
\end{equation}
wherein $\phi''=\phi''\left(\phi\right)$, i.e.\ $\sin\left(\phi''+\phi\right)=n' \sin\left(\phi\right)$. Eq.\ \ref{eq:CZA} describes a circle, see dashed line in Fig.\ \ref{Fig_Exp}(b),(c). However, it turns out that only a segment of the circle is attainable by the exiting rays due to the occurrence of total internal reflection. The solid black line in Fig.\ \ref{Fig_Exp}(c) shows this limit. The critical internal angle of incidence may be found from $\phi_{\rm TIR}=\arcsin\left(1/n'\right)$, such that $\phi'=\pi/2$ marks the onset of total internal reflection. Herein $n'$ is a function of $e'$ which is a function of $e$. One finds\cite{Bravais1847,Wegener}%p. 333 (Note VI, 5e Cas, (49))?
\begin{equation}
\phi_{\rm TIR}=\arccos\left(\sqrt{n_0^2-1}/\cos\left(e\right)\right)\label{Eq:phiTIR},
\end{equation}
which translates into a corresponding azimuthal limit $\phi''_{\rm TIR}=\pi/2-\phi_{\rm TIR}$ of the (projected artificial) CZA.

A similar reasoning leads to the existence of a critical elevation angle $e_{\rm TIR}$ above which the internal second refraction becomes a total internal reflection, $e''_c=0$, Eq.\ \eqref{Eq:eppc}, even for $\phi=0$ where $n'$ is lowest. Equivalently, one may set $\phi_{\rm TIR}\rightarrow 0$ and solve Eq.\ \eqref{Eq:phiTIR} for $e$ to arrive at:\cite{Bravais1847,Humphreys,Wegener,McDowell1979}
\begin{equation}
e_{\rm TIR}=\arccos\left(\sqrt{n_0^2-1}\right)\label{Eq:eTIR}.
\end{equation}
Eq.\ \eqref{Eq:eTIR} shows that at around $e_{\rm TIR}=28^{\circ}$ even the last glimpse of the red ($n_0\left(\textnormal{red}\right)=1.332$, i.e.\ less refracted than blue $n_0\left(\textnormal{blue}\right)=1.341$) part of the artificial (water) CZA disappears. For ice, taking $n_0=1.31$, the corresponding critical solar elevation above which this halo can no longer be observed is $32^{\circ}$.\cite{Bravais1847,Wegener,Humphreys,Ticker,McDowell1979} Eq.\ \eqref{Eq:eTIR} also shows that any material with $n_0>\sqrt{2}$, i.e.\ glass, will not produce a CZA (nor a CHA).\cite{Tammer1998,Vollmer2014,Selmke2015,Borchard2015} For this reason alone, and in order to not have to construct a water-filled hexagonal prism, it is nice to have a simple analog demonstration experiment to overcome this practical limitation. The full azimuthal width of the CZA is $\Delta \phi''_{\rm CZA}=2\phi''_{\rm TIR}$ and is an increasing function of the elevation, starting from $125^{\circ}$ and approaching a half-circle, i.e.\ $180^{\circ}$, for $e\rightarrow e_{\rm TIR}$. In this limit, Eqs.\ \eqref{Eq:phiTIR} and \eqref{Eq:eTIR} show that light emerges only from a small section around $\phi=0$, where the effective index of refraction diverges $n'\left(e_{\rm TIR}\right)\rightarrow \infty$ whereby the exiting refraction deviates rays at a right angle $\phi'\approx \pi/2$ towards the left and the right. %Bravais: p.96, Humphreys. p. 512

The complementary angle $e_c''$ of the final exit ray's inclination, Eq.\ \eqref{Eq:eppc}, corresponds to the angular distance to the azimuth of the natural CZA halo phenomenon.\cite{CZAnat,Wolfram,note46} This angular distance is independent of the azimuth $\phi''$ (or $\phi$), such that the natural CZA appears as a true circle around the zenith (Fig.\ \ref{Fig_Exp}(a)),\cite{Bravais1847,Humphreys,Wegener} just as the artificial CZA is a circle in the $xy$-plane (Fig.\ \ref{Fig_Exp}(b)). One may also observe, both in natural displays of the phenomenon as well as in the experiment, that the angular width of the visible spectrum, i.e.\ the chromatic angular dispersion $\Delta e''_c=e''_c\left(\textnormal{red}\right)-e''_c\left(\textnormal{blue}\right)$, remains roughly constant at $1.6^{\circ}$ (which may be compared to the dispersion $\sim 1.2^{\circ}$ of the primary rainbow). Only at very small inclinations $e$ a broadening is observed\cite{Wegener,McDowell1979} before the shrinking CZA eventually disappears as it converges towards the zenith (or the point below the cylinder), cf.\ Fig.\ \ref{Fig_Exp}(a),(c).

\section{Arc Intensity\label{Appendix:Int}}
\begin{figure}[bt]
\begin{center}\includegraphics [width=1.0\columnwidth]{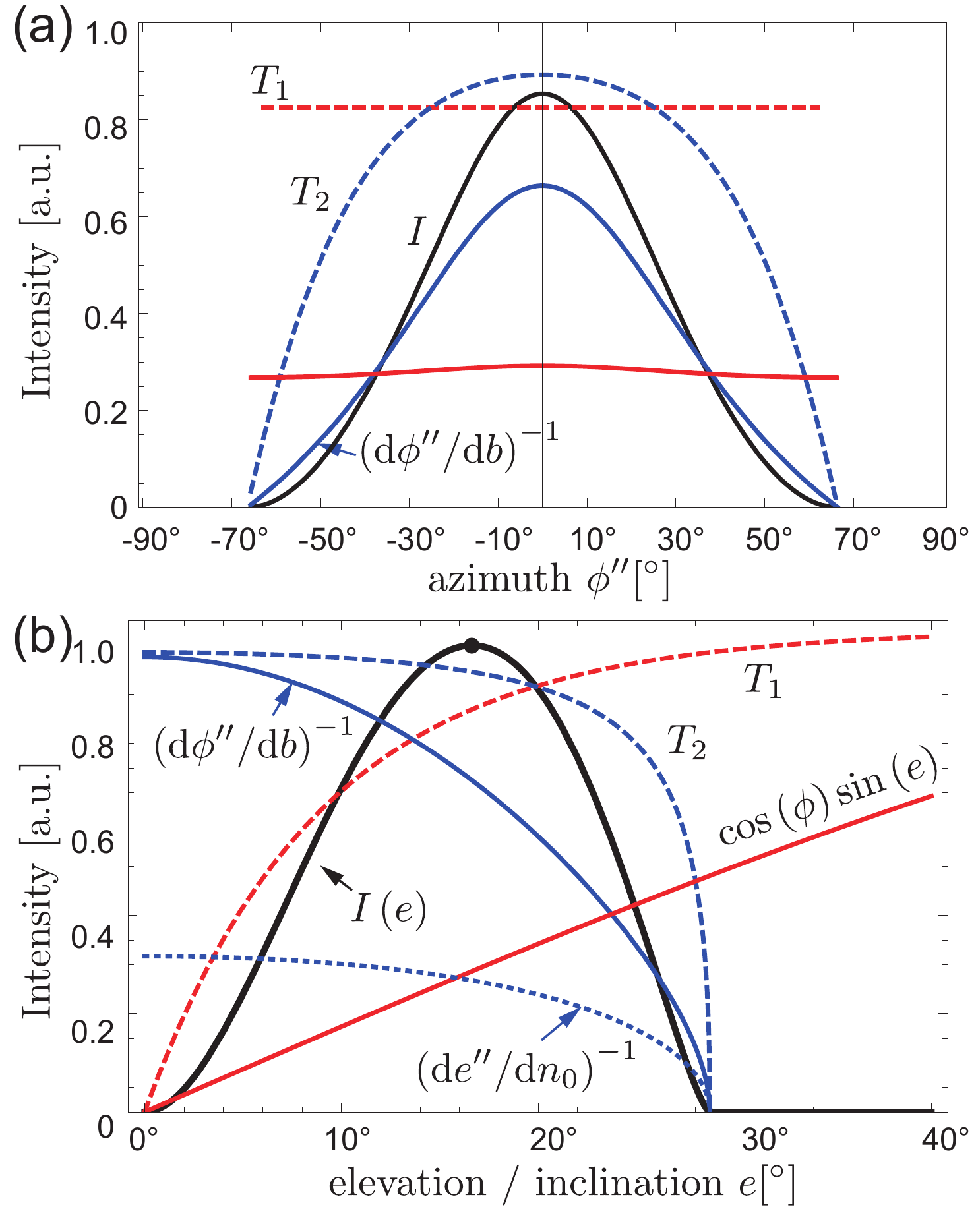}\end{center}
\caption{The CZA intensity distribution along \textbf{(a)} the azimuth coordinate $\phi''$ at $e=17^{\circ}$. Only for $\phi'' \in [-\phi''_{\rm TIR},+\phi''_{\rm TIR}]$ is $I\left(\phi''\right)>0$. \textbf{(b)} CZA intensity at $\phi''=\phi=0$ for different elevation (inclination) angles $e$. Only for $e\in [0,e_{\rm TIR}]$ is $I\left(e\right)>0$. The refractive index of water was used here, $1.33$. The picture for ice is very similar.\label{Fig_Intensity}}
\end{figure}

Without treating the situation in full detail, a description of the approximate intensity along the azimuthal coordinate $\phi''$ requires several key factors to be considered:\cite{McDowell1979}
\begin{itemize}
\item $T_1$, transmission at the first interface,
\item $T_2$, transmission at the second interface,
\item $A\propto \cos\left(\phi\right) \sin\left(e\right)\cot\left(e'\right)$, a geometric factor\cite{crosssection},
\item $\left(\mathrm{d}\phi''/\mathrm{d}b\right)^{-1}$, $b=\sin\left(\phi\right)$, i.e.\ a ray bundleling factor,
\item $\left(\mathrm{d}e''/\mathrm{d}n_0\right)^{-1}$, a chromatic dispersion factor.
\item $I_{AM}$, atmospheric attenuation (natural halo only)
\end{itemize}
The cross-sectional factor $A$ takes the projected surface of the top face into account which admits refraction along the ray path described by the internal angle $\phi$, see dashed area in Fig.\ \ref{Fig_Geometry}(c).\cite{crosssection} The ray bundleing factor is a caustic intensity factor for the cylinder experiment and a fake caustic intensity factor for the natural counterpart.\cite{Koennen1983} Its concept is similar to that of the rainbow caustic,\cite{Berry2015} and accounts for the predominance of certain deflection directions as outcomes of the refraction of an incident parallel bundle of rays. The fact that more rays experience a small azimuthal in-plane refraction is then captured by this factor peaking around $\phi''\approx 0$. However, in this case no divergence of the intensity appears, i.e.\ this factor remains finite at all times. The chromatic dispersion factor accounts for the changing width $\Delta e''_c\approx \left(\mathrm{d}e''/\mathrm{d}n_0\right) \times \left(n_0\left(\textnormal{blue}\right)-n_0\left(\textnormal{red}\right)\right)$ of the arc, such that its apparent brightness is $\propto 1/\Delta e''_c$. Overall then, 
\begin{equation}
I\left(\phi'',e\right)\propto T_1\cdot T_2 \cdot A \cdot \left(\mathrm{d}\phi''/\mathrm{d}b\right)^{-1} \cdot \left(\mathrm{d}e''/\mathrm{d}n_0\right)^{-1},\label{Eq:Int}
\end{equation}
cf.\ Fig.\ \ref{Fig_Intensity}. This expression quantifies the observed azimuthal decay in intensity away from the forward direction and towards zero for $\phi''\rightarrow \phi''_{\rm TIR}$ due to the second interface's transmission going smoothly to zero as the limit of total internal reflection is reached. The transmission coefficient $T_2\left(\varphi\right)$ (blue dashed line in Fig.\ \ref{Fig_Intensity}(a)) upon the second refraction depends on the actual angle $\varphi$ to the normal, cf.\ Fig.\ \ref{Fig_Geometry}(d), which is given by $\varphi=\arccos\left(\cos\left(\phi\right)\cos\left(e\right)/n_0\right)$.\cite{T2angle} The parametric curves in Fig.\ \ref{Fig_Exp}(c) have been shaded according to this intensity function and match the appearance of experimentally observed CZAs projections and natural halo displays well.

Plotting the intensity in the forward direction at the position of maximum CZA intensity (at $\phi''=\phi=0$) as a function of the elevation (inclination), $I\left(e\right)$, one observes a wide peak, see thick black line in Fig.\ \ref{Fig_Intensity}(b). This means that for some solar elevations (or light source inclinations) the (artificial) CZA is brighter as compared to others. For water and ice (including the atmospheric attenuation), this peak occurs at around $17^{\circ}$ and $20^{\circ}$, respectively. For ice then, this value corresponds to the elevation at which the CZA is best observable, see for instance Refs.\ \cite{Humphreys,Wegener,McDowell1979,atmosHP}. For water, this is the light source inclination for which the experiment produces the brightest CZA projection. Both $T_1$ (red dashed line) and the geometric factor (red solid line) describe the decay towards $e\rightarrow 0$. For the natural halo, also the atmospheric attenuation affects the decay when the sun is low and dim\cite{AirMass}. At the other end of the curve, $T_2$ (blue dashed line) along with the ray bundleing factor (blue solid line) describe the decay towards $e\rightarrow e_{\rm TIR}$. The decrease of the ray bundleling factor may be understood as follows: The inclination angle $e'$ increases for increasing elevation angles $e$, such that the effective index of refraction $n'$, Eq.\ \eqref{eq:neff}, increases as well (eventually diverging as we have seen). This in turn causes the exit rays which are deflected by $\phi''$ to sweep more rapidly across the forward ($\phi''=0$) direction as the impact parameter $b$ crosses the symmetry axis. Accordingly then, the bundleling of rays in the forward direction is reduced as $e$ increases.

\section{Artificial circumhorizontal arc}
\begin{figure}[t]
\begin{center}\includegraphics [width=1.0\columnwidth]{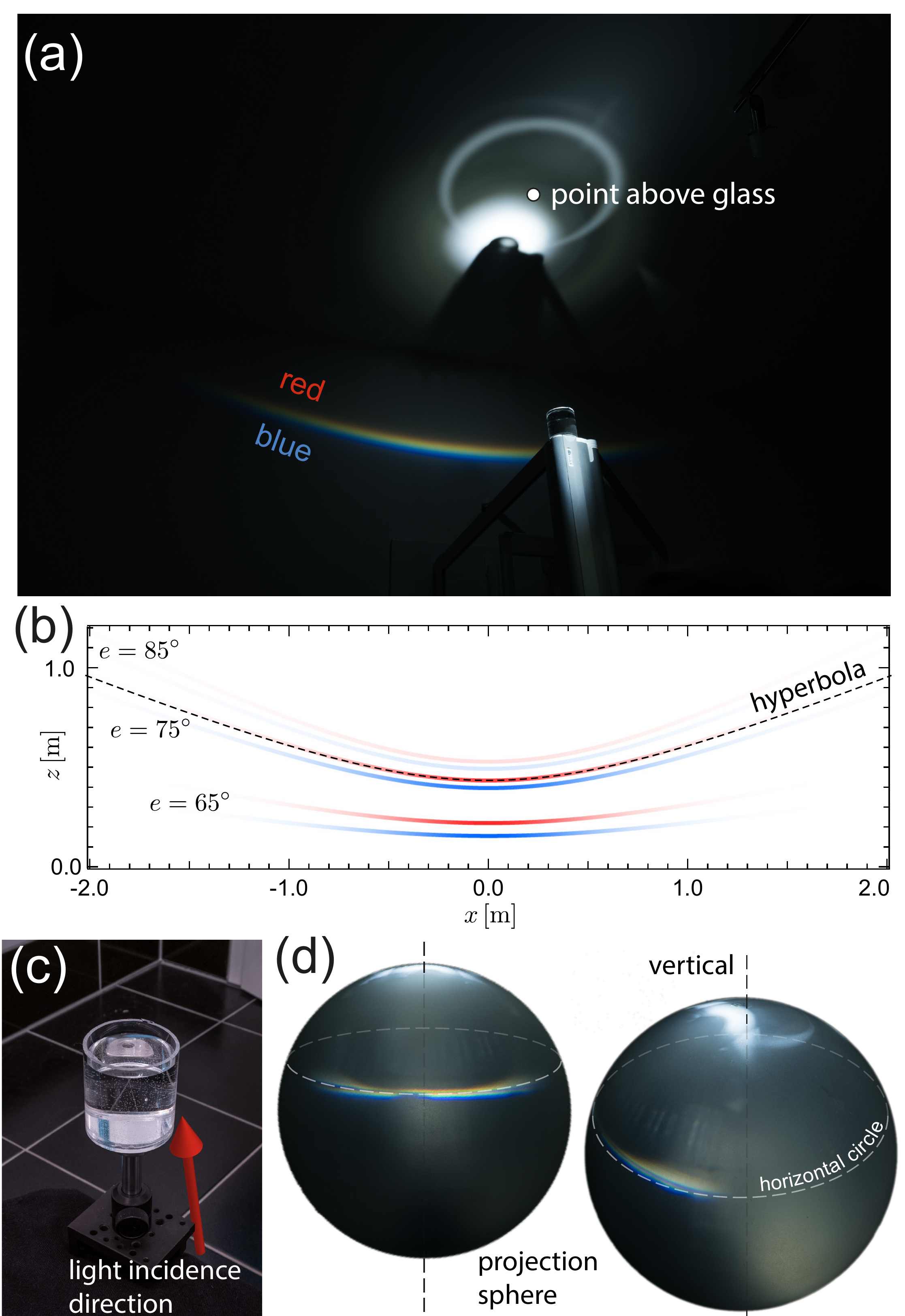}\end{center}
\caption{\textbf{(a)} Artificial circumhorizontal arc (CHA) projection. The circle at the ceiling, crossing the shadow of the water-filled cylinder, corresponds to the natural parhelic circle halo (the external reflection contribution). \textbf{(b)} Plot of the CHA curve, Eq.\ \ref{Eq:CHA}, for elevations $e=\left\{65^{\circ},75^{\circ},85^{\circ}\right\}$ and for red and blue color each ($l=1\,\rm m$). \textbf{(c)} The acrylic cylinder as placed in the center of a spherical projection screen.\cite{SelmkeSelmke2015} \textbf{(d)} Two views showing the CHA projection onto the spherical screen.\label{Fig_CHA}}
\end{figure}

We now turn to the artificial CHA, see Fig.\ \ref{Fig_CHA}(a), for which the first refraction is the side wall of the cylinder in the experiment, see Fig.\ \ref{Fig_Geometry}(e)-(g). Accordingly, it is this refraction which must be treated according to the inclined skew-ray theory of Bravais. Again, we decompose the problem into two parts: The material's index of refraction $n_0$ directly determines the change in the inclination angle, $\sin\left(e\right)=n_0\sin\left(e'\right)$, and the in-plane refraction (i.e.\ as seen from above) follows $\sin\left(\phi\right)=n'\sin\left(\phi'\right)$, with the effective index of refraction being\cite{Bravais1847,Wegener,Humphreys,Ticker,Koennen1983,Selmke2015}
\begin{equation}
n'=\sqrt{\frac{n_0^2 - \sin^2\left(e\right)}{\cos^2\left(e\right)}}.\label{eq:neffOrig}
\end{equation}
The second refraction at the top water-air interface only changes the inclination towards the horizontal, $n_0\sin\left(e'_c\right)=\sin\left(e''_c\right)$, and no in-plane refraction takes place since the refracting interfaces' normal is vertical. One finds $\cos\left(e''\right)^2=n_0^2-\sin^2\left(e\right)$, which is the analog of Eq.\ \eqref{Eq:eppc}. Referring to the experiment's setup and coordinates as defined in Fig.\ \ref{Fig_Geometry}(e)-(g), the curve $\left(x,z\right)$ described by the artificial CHA on the vertical wall is parametrized by the angle $\phi$ as:
\begin{equation}
\left(\begin{array}{c}x\\z\end{array}\right)=\frac{l}{\cos\left(\phi''\right)}\left(\begin{array}{c}\sin\left(\phi''\right)\\\tan\left(\phi''\right)\end{array}\right),\label{Eq:CHA}
\end{equation}
%%Bravais: p. 233 (Note VI, 6e Cas, (51))? l=n0, alpha=A (prisms' apex angle), H=e (elevation), h=e''
where $\phi''=\phi-\phi'$. Since $z^2 = \left(x^2+l^2\right)\tan^2\left(e''\right)$, and $\tan^2\left(e''\right)$ is a constant independent of $x$,\cite{hyperbola} the artificial projected CHA for each inclination $e$ is a hyperbola in the $xz$-plane, see Fig.\ \ref{Fig_CHA}(b). If the projection were onto a sphere, the natural CHA halo's geometry of a circle segment parallel to the horizon (i.e. at constant elevation) were to be recovered, cf.\ Ref.\ \cite{SelmkeSelmke2015}. The angle to the surface normal which determines the transmission coefficient $T_1\left(\varphi\right)$ is $\varphi=\arccos\left(\cos\left(\phi\right)\cos\left(e\right)\right)$, see Fig.\ \ref{Fig_Geometry}(h). The transmission coefficient $T_2\left(e'_c\right)$ corresponds to the second refraction changing the elevation only. The intensity may then be analyzed along the same lines as for the CZA, but with the cross-sectional factor being $\cos\left(\phi\right)\cos\left(e\right)$.\cite{CHAint} Again, this approach was used to set the transparency of the curves in Fig.\ \ref{Fig_CHA}(b). We find that the CHA is brightest at around $e=69^{\circ}$ in the experiment and $65^{\circ}$ for the natural halo, whereas azimuthally it decays to zero for $\phi\rightarrow \pi/2$ (grazing incidence) where accordingly $\phi''=\pi/2-\arcsin\left(1/n'\right)$.\cite{Bravais1847,Wegener} The corresponding full azimuthal width $\Delta\phi''_{\rm CHA}$ of twice that value therefore ranges from $125^{\circ}\rightarrow 180^{\circ}$ in the experiment and $116^{\circ}\rightarrow 180^{\circ}$ for the natural ice halo phenomenon. The half-circle limit is reached in the reverse situation as compared to the CZA, i.e.\ here for grazing incidence with $e=\pi/2$ (as compared to grazing exit $e''_c=\pi/2$ for the CZA). The natural CHA only forms when the inclination is larger than $\pi/2-e_{\rm TIR}=58^{\circ}$ (ice\cite{Wegener}) and $62^{\circ}$ (water), where the critical inclination as determined for the CZA, Eq.\ \eqref{Eq:eTIR}, can be reused. The halo rises in altitude (elevation) as the sun approaches the zenith ($90^{\circ}$ angle of incidence onto the side face), where it will be $e''=32^{\circ}$.\cite{Bravais1847,Ticker} The experimental CHA reproduces these behaviors closely.

%\vspace{0.5cm}
\section{Conclusion \& Outlook}
A thorough analysis of a very simple experiment, namely a glass filled with water illuminated under various directions of incidence, provides a rich phenomenology. We have shown that the emerging projections of light closely correspond to two natural atmospheric ice halos: the circumzenithal and the circumhorizontal arcs. The general angular characteristics derived and validated by the experiment also apply to their natural counterparts. This demonstration experiment may complement more complex ones based on spinning glass crystals\cite{Tammer1998,Greenler2003,Vollmer2014,Borchard2015,Selmke2015,SelmkeSelmke2015}. It also produces a purer spectrum as compared to a rainbow demonstration since, similar to the action of a prism, ideally no color-overlap occurs. The experiment may thus be used as a halo alternative to rainbow demonstrations\cite{Casini2012} or as an illustrative example of skew-ray refraction.

We end with an outlook on similar experiments: One may use a water-filled martini / cosmopolitain cocktail glass as a refracting cone, cf.\ Fig.\ \ref{Fig_ParryAvg}. By the same idea that led us to the CZA analogy, one may confirm (see Appendix \ref{AppendixParry}) that the average geometry of light entering through the top air-water interface and leaving through the lateral cone surface results in an analogy to Parry's halo.\cite{Wegener,Greenler1980,TapeBook} Using a cocktail glass, one may find in addition an artificial parhelic circle due to external reflections from the stem of the glass as well as artificial heliac arcs\cite{TapeBook} due to external reflections by the conical surface. Likely, many more halo counterparts may be realized along the lines presented here.

%We may use the same glass an put something in its center to emulate the opaque core that Huygens assumed to cause the appearance of parhelia left and right from the sun. One may show that the appropriate ratio of the outer glass diameter to the inner opaque should be exactly $1/2$ to match the corresponding parhelia angle.\cite{PHHuygens} %p.234-236 "Compleat Opticks", also: TapeJarmo2006

%
\begin{figure}[b]
\begin{center}\includegraphics [width=1.0\columnwidth]{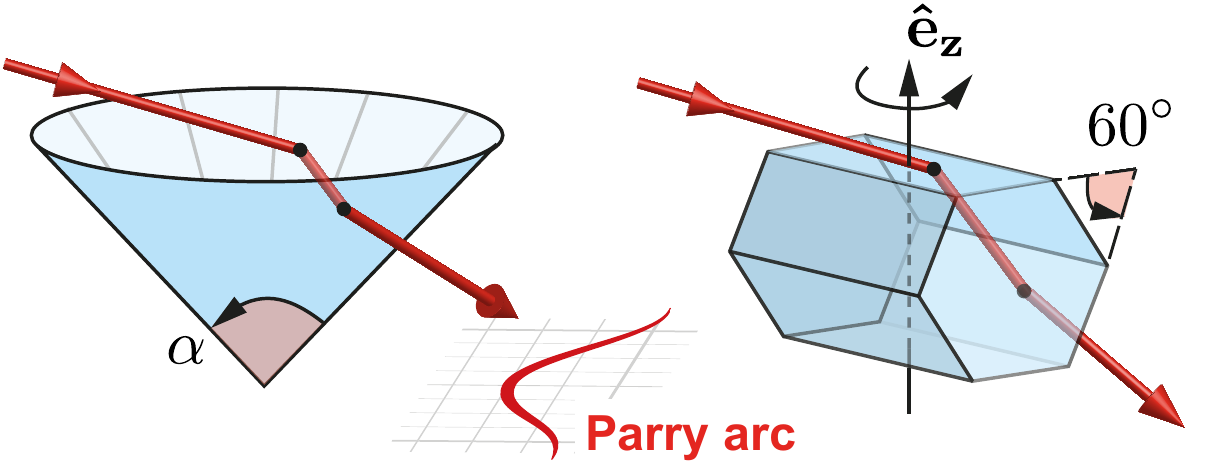}\end{center}
\caption{Analogy (in the sense of Fig.\ \ref{Fig_Corresp}) of the parallel light refraction by a cone (e.g. a filled cocktail / martini glass) and the ray path responsible for the upper suncave Parry arc.\cite{Wegener,Greenler1980,TapeBook} While the cocktail glass typically has an apex angle of $\alpha \sim 70^{\circ}$ instead of the $60^{\circ}$ required for a perfect analogy, the resulting pattern is very similar.\label{Fig_ParryAvg}}
\end{figure}

\appendix
\section{Colors}
Perceived colors were computed as follows: First, the tristimulus values (trichromatic color space coordinates) $\left\{X,Y,Z\right\}$ are computed via the CIE standard calorimetric observer color matching functions $\left\{\overline{x},\overline{y},\overline{z}\right\}$ and from the spectral radiance $L$, e.g.\ $X=\int_0^\infty L\left(\lambda\right)\overline{x}\left(\lambda\right)\mathrm{d}\lambda$. We assumed a linear relation between this quantity and the spectral power density of a given light source. For a single wavelength, a narrow (FWHM $\sim 1\,\rm nm$) Gaussian line spectrum was chosen. Then, a linear transformation (a matrix product) converts these to linear RGB values, $\left(R_L,G_L,B_L\right)=M\cdot \left(X,Y,Z\right)$, which subsequently are converted to the common sRGB color space values through a non-linear transformation. Finally, the values are normalized to give a valid sRGB color. The required data can be found online at \url{http://www.cie.co.at} and \url{https://en.wikipedia.org/wiki/SRGB}. The hereby obtained sRGB color values $\left\{R_s,G_s,B_s\right\}$ for each wavelength were multiplied by an intensity factor $I/I_0$ to recreate the azimuthal intensity decay of the correspondingly colored CZH segments corresponding to $n_0\left(\lambda\right)$.

\begin{figure*}[htb]
\begin{center}\includegraphics [width=1.0\textwidth]{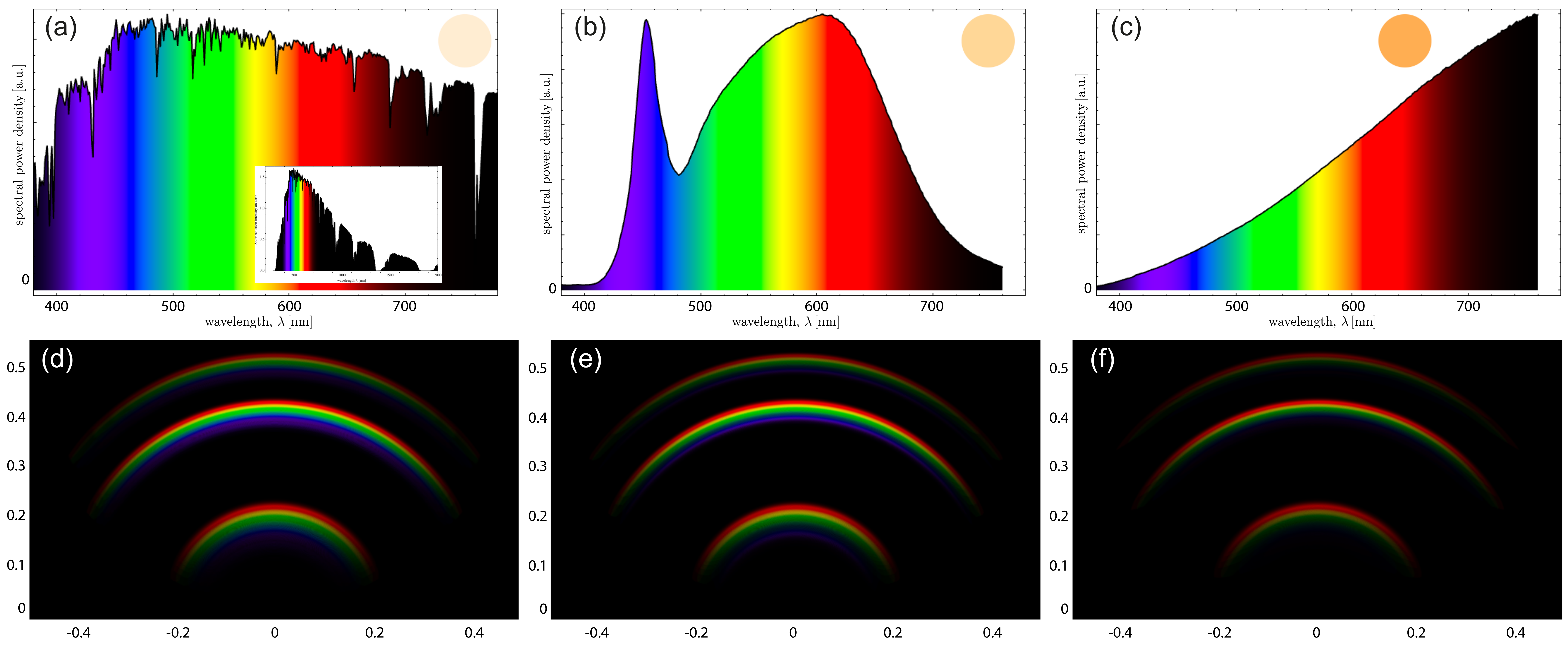}\end{center}
\caption{Light source spectral power densities for \textbf{(a)} sun light \textbf{(b)} LED light \textbf{(c)} incandescent light bulb. The colors of the sources are given shown in the spectrum's insets. \textbf{(d)-(f)} Corresponding perceived colors of the artificial CZAs in the $xy$-plane.\label{Fig_CZAColor}}
\end{figure*}

Corresponding computation assuming a solar spectrum, a LED light source spectrum and a n incandescent light bulb spectrum along with the dispersion $n_0\left(\lambda\right)$ of water are shown in Fig.\ \ref{Fig_CZAColor} for different elevations. They may be compared to displays computed with dedicated halo simulation software.\cite{HaloSim} Note that in the experiment colors may superimpose (and green disappear) if the projection distance $l$ is too small.

%\begin{equation}
%\left(\begin{array}{c}R_L\\G_L\\B_L\end{array}\right) = \left(\begin{array}{ccc}3.2405 & -1.5372 & -0.4985\\ -0.9693 & 1.8760 & 0.04156\\ 0.0556 & -0.2040 & 1.0573\end{array}\right) \cdot \left(\begin{array}{c}X\\ Y\\ Z\end{array}\right)
%\end{equation}

\section{Artificial suncave Parry arc\label{AppendixParry}}
We keep the notation of Fig.\ \ref{Fig_Geometry}(a), where we imagine the cylinder being replaced by the cone of Fig.\ \ref{Fig_ParryAvg}, and acknowledge that for the second refraction (the first being analogous to the CZA) the normal vector is now $\mathbf{\hat{n}}=\left(-\cos\left(\alpha/2\right) \sin\left(\phi\right), -\cos\left(\alpha/2\right) \cos\left(\phi\right), \sin\left(\alpha/2\right)\right)$ for the cone's lateral surface. The incidence vector $\mathbf{\hat{i}}=\left(0,\sin\left(e'_c\right),-\cos\left(e'_c\right)\right)$ makes an angle $\varphi=\arccos(-\mathbf{\hat{i}}\cdot \mathbf{\hat{n}})$ to this normal. The refracted angle is then determined by $n_0 \sin\left(\varphi\right)=\sin\left(\varphi'\right)$. To get the refracted ray, one may rotate the external unit vector $-\mathbf{\hat{n}}$ according to the right hand rule by $\varphi'$ about an axis perpendicular to the plane of incidence, defined by the unit vector $\mathbf{\hat{k}}=\mathbf{\hat{i}}\times \mathbf{\hat{n}}/\sin\left(\varphi\right)$. This can be done via Rodrigues' rotation formula, yielding the exiting ray's unit direction vector $\mathbf{\hat{r}}=\mathbf{\hat{n}}\left(n_0\cos\left(\varphi\right)-\cos\left(\varphi'\right)\right) + \mathbf{\hat{i}} n_0$. The line described by this unit direction vector and starting at a height $l$ above the origin (at the position of the glass) intersects the horizontal $xy$-(projection-)plane at the point $\left(x,y,0\right)=l \mathbf{\hat{e}_z} -l \mathbf{\hat{r}}/\left(\mathbf{\hat{n}}\cdot \mathbf{\hat{r}}\right)$. This is the artificial projected suncave Parry arc parametrized by $\phi$. The complementary angle to the elevation is $e''_c=\arccos\left(\mathbf{-\hat{e}_z}\cdot \mathbf{\hat{r}}\right)$ when $\mathbf{\hat{r}}_y>0$, and negative that value when $\mathbf{\hat{r}}_y<0$, signifying a Parry arc beyond the zenith with $e''>\pi/2$ (in front of the cocktail glass when viewed from the direction of incident light). The azimuths are $\phi''=\pm \arccos\left(\left(0,1,0\right)\cdot \left(\mathbf{\hat{r}}_x,\mathbf{\hat{r}}_y,0\right)/|\left(\mathbf{\hat{r}}_x,\mathbf{\hat{r}}_y,0\right) |\right)$ for $\mathbf{\hat{r}}_x\lessgtr 0$. The arc appears concave towards the shadow of the glass (corresponding to the sun position and being suncave as its natural counterpart), and the angular distance to this shadow, $\Delta_S=\arccos(\left(0,\cos\left(e\right),-\sin\left(e\right)\right)\cdot \mathbf{\hat{r}})$, corresponds to the distance to the sun for the natural halo. The hereby obtained angular coordinates agree with e.g.\ Ref \cite{Wegener}. Alternatively then, the projected artificial Parry arc curve on the floor may be computed via Eq.\ \eqref{eq:CZA} if the following replacements via the variables of Wegener's chapter 12\cite{Wegener} are made: $e''_c\rightarrow \pi/2-h_\sigma$ and $\phi''\rightarrow \delta$ and Wegener's $\varphi$ being the parametrization. Using glasses with smaller apex angles $\alpha$ (walls becoming increasingly vertical / cylinder-like), the arc inverts from suncave to sunvex and approaches the CZA in the limit $\alpha\rightarrow 0$. 

%\clearpage

\renewcommand{\arraystretch}{2}
\renewcommand{\tabcolsep}{9pt}

%\bibliographystyle{unsrt}
%\bibliography{PhysTeacher}

\begin{thebibliography}{19}
\bibitem{Distance}
The distance must be much larger than the focal distance of the curved cylinder interface, $f=R n'/\left(n'-1\right)$, where $R$ is the cylinder's radius. Practically, $\gtrsim 10 R$ is fine.
\bibitem{CZAnat}
The ratio $r=D/w=0.347$, where $D$ is the CZA diameter and $w$ the image width, may be used to extract the angular distance to the zenith, $e''_c=\arctan\left(r\cdot X/2f\right)=16^{\circ}$. This value is in agreement with Eq.\ \eqref{Eq:eppc}, $e=27^{\circ}$\cite{Wolfram} and $n=1.31$. $f=14\,\rm mm$ was the focal length of the rectilinear projection lens and $X=23.6\,\rm mm$ the APS- C sensors' $x$-dimension.
\bibitem{CHAint}
To include the effect of the reduced projection intensity on the wall due to the inclination of the rays, one may further include a factor $\sim \cos\left(\phi''\right)$. We choose to ignore it here to be in line with the natural CHA intensity.
\end{thebibliography}

\end{document}